\documentclass[a4paper,11pt]{article}
\usepackage{jcappub} % for details on the use of the package, please see the JINST-author-manual
\usepackage{lineno}
\usepackage{xcolor}

\title{Fast Identification of Transients: Applying Expectation Maximization to Neutrino Data}

\author[a, b]{M. Karl}
\author[a]{P. Eller}

\affiliation[a]{Technische Universit{\"a}t M{\"u}nchen, TUM School of Natural Sciences, Physics Department, 
James-Frank-Str. 1, D-85748 Garching bei M{\"u}nchen, Germany}
\affiliation[b]{European Southern Observatory, Karl-Schwarzschild-Str. 
2, D-85748 Garching bei M\"unchen, Germany\\}
\emailAdd{martina.karl@tum.de, philipp.eller@tum.de}

\abstract{
We present a novel method for identifying transients suitable for both strong signal-dominated and background-dominated objects. By employing the unsupervised machine learning algorithm known as Expectation Maximization, we achieve computing time reductions of over $10^4$ on a single CPU compared to conventional brute-force methods. Furthermore, this approach can be readily extended to analyze multiple flares. We illustrate the algorithm's application by fitting the IceCube neutrino flare of TXS~0506+056.
} 

\begin{document}
\maketitle
\flushbottom

\section{Introduction}\label{sec:intro}

Transient objects emit variable signals over time. Many astrophysical objects are transient, including the most powerful objects in our universe, such as blazars \cite{Padovani_2017} (a subclass of active galactic nuclei), gamma-ray bursts \cite{annurev:/content/journals/10.1146/annurev.aa.28.090190.002153}, or supernovae. In many cases, investigating the variable nature of these sources is supported by bright flares and clearly identifiable high-emission states. Various methods for identifying transient signals have been devised for Astronomy and other fields, such as, for example, Bayesian Blocks \cite{Scargle_2013}. For signal-dominated emission, Bayesian Blocks provides reliable identifications of the flaring periods (see, for example, ref. \cite{10.1093/mnras/stad2724}). 
When the time of a flare is known from an observation in a different channel or messenger, data can be searched even for sub-threshold transient emissions including this information. This was, for example, done in the multi-messenger observation of the blazar TXS~0506+056 following the IceCube alert 170922A \cite{IceCube:2018dnn}, where the time of the alert defined the time window of interest for follow-up surveys. Similar approaches can be used for gamma-ray bursts or supernovae.
%There, the gamma-ray flare of TXS~0506+056 defined the time-window of interest.  when looking for neutrino emission coincident with the gamma-ray flare of  As was the case, for example, in the multi-messenger observation of the blazer TXS~0506+056 following the IceCube alert 170922A \cite{IceCube:2018dnn} or the binary neutron star merger \cite{LIGOScientific:2017ync}.

However, identifying time-variable signals can be challenging if the source's active time period is unknown and the signal is dominated by background. This scenario was, for example, encountered in the case of searching IceCube's archival data for transient emissions from the source location of TXS~0506+056 prior to the IceCube alert 170922A \cite{2018Sci...361..147I}.
Since Bayesian Blocks aims for bins with equal flux content, identifying small signal fluxes compared to the background flux is challenging. We present an approach that identifies both strong emission states and weak background-dominated emissions. Such background-dominated signals are often found in neutrino astronomy, an emerging field studying transient sources. The difference in our work compared to Bayesian Blocks and many other approaches is that we assume we know the likelihood function describing the background and the signal process, whereas others aim to make minimal assumptions. This makes our search more sensitive to signals well described by our likelihood. However, whenever possible, it is, of course, favorable not to have to include specific signal assumptions if it is not necessary for flare identification. Hence, strong signal-dominated flares, such as gamma-ray bursts or supernovae, will profit from an approach without any prior assumptions. In this work, we focus on identifying flares that are not easily distinguishable from the background. In general, the same method can, of course, also identify strong flares.

Conventional methods for identifying background-dominated flares as in refs. \cite{2018Sci...361..147I, abbasi2021search, aartsen2015searches, 2021ApJ...911...67A} calculate signal weights for each event and apply brute force scans where all possible intervals of events with weights exceeding a certain threshold are either evaluated as possible starting and end times of a neutrino flare or used as a seed for subsequent optimization of parameters. In ref. \cite{2018Sci...361..147I}, the threshold was very small, i.e., all events better described by the signal hypothesis than the background hypothesis were identified as possible starting and endpoints. Adopting this approach leads to immense computing times. Especially since the available data of the IceCube Neutrino Observatory (or neutrino data in general with new neutrino telescopes in construction) keeps increasing, running expensive searches on large sections of the sky on 14+ years of neutrino data becomes more and more computationally infeasible.

Increasing the signal weight threshold is one way to reduce the computing time, as was done, for example, in ref. \cite{abbasi2021search}. 
However, this also means evaluating reduced information and potentially favoring specific model parameters since the assumed model parameters enter the signal weight calculation. 
References \cite{2022icrc.confE.940K, dissertation_martina} investigated new approaches, and we conducted a first search applying the method presented in this work (an unsupervised machine learning algorithm) on IceCube data in Ref.~\cite{martina_alert:2023icrc, alert_followup_paper, dissertation_martina}. In the following, we describe a general approach applicable to any model and data in Sec.~\ref{sec:flares} and show the application for identifying the neutrino flare of TXS~0506+056 in Sec.~\ref{sec:TXS}. For this, we analyze published data of through-going muon-tracks \cite{IceCube:2021xar} of the IceCube Neutrino Observatory using the open source framework SkyLLH\footnote{\url{https://github.com/icecube/skyllh}} \cite{skyllh:2023icrc}.

\section{Identification of Transient Signals}\label{sec:flares}

The example in Fig.~\ref{fig:time_histogram} shows a uniform distribution of $\approx 21000$ background events over eleven years and five signal events from a transient emission around $t=57000$\,MJD. We cannot recognize the signal by eye, and trying to identify a signal by just studying the frequency of events detected is futile.
We can, however, try to incorporate as much knowledge on the background events and the signal events we are looking for as possible. The best separation between signal and background is achieved by their likelihood ratio, as stated in the famous ``Neyman-Pearson'' lemma \citep{doi:10.1098/rsta.1933.0009}. This comes at the cost of defining a likelihood for the signal and background process, respectively, but will render our search much more sensitive. We assume here the following for our background and signal-generating processes:

\begin{figure}
    \centering
    \includegraphics[width=\textwidth]{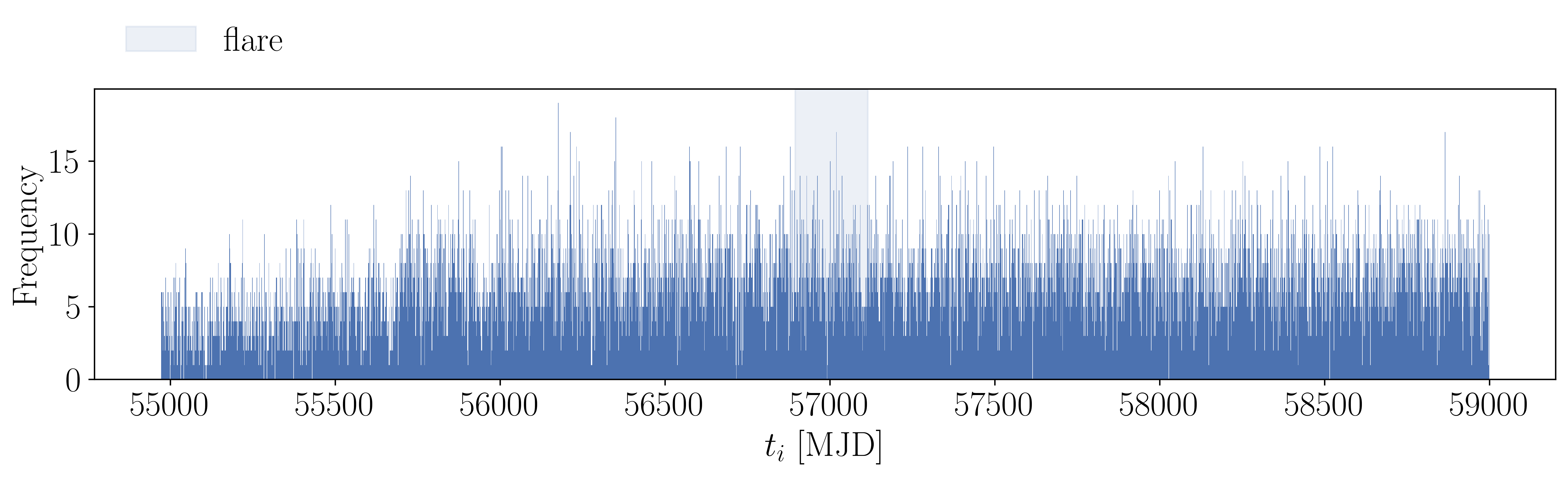}
    \caption{Histogram (or lightcurve) of detected event arrival times. There is a flare of five signal events in the highlighted time range.}
    \label{fig:time_histogram}
\end{figure}

\begin{itemize}
    \item \textbf{Background Process}: The data originates from some time-invariant, continuous process, i.e. flat as a function of time. We assume that we know the energy distribution of the events and their dependence on the location $\vec{x}$. This information can be inferred by studying enough background data. %contains atmospheric neutrinos, atmospheric muons, and diffuse astrophysical neutrinos. We assume a uniform flux in time and right ascension\footnote{Due to IceCube's unique position at the Sout Pole, we expect a uniform spatial background distribution in right ascension for integration times longer than a day.}.\\
    \item \textbf{Signal Process}:  We assume signal events to originate from a spatially defined source and cluster around the source position $\vec{x}_{\rm{S}}$. Furthermore, we can make some assumptions on the energy distribution of signal events. In the following examples, we assume a point-like source since most transient objects (active galactic nuclei, pulsars, supernova remnants, etc) have extensions not resolvable with neutrino \citep{IceCube:2021xar} or gamma-ray data\footnote{\url{https://fermi.gsfc.nasa.gov/ssc/data/analysis/documentation/Cicerone/Cicerone_LAT_IRFs/IRF_PSF.html}}. We furthermore assume an energy spectrum following a power law: $\frac{d \phi}{d E_{\nu}} \propto E_{\nu}^{-\gamma}$. Furthermore, the signal component is transient, i.e. time-dependent, and follows either a Gaussian time profile (with mean $\mu_{\rm{T}}$ and width $\sigma_{\rm{T}}$) or a box profile with constant emission during a flare window with starting and end time ($t_{\rm{start}}$, $t_{\rm{end}}$).\\
\end{itemize}

With this information at hand, we can define probability density functions (pdfs) for the signal process $\mathcal{S}$ and the background process $\mathcal{B}$, respectively, as described in refs. \cite{BRAUN2008299, skyllh:2023icrc}. We divide these pdfs into spatial, energy, and temporal parts. For generalization, we refer to the temporal parameters as $t_a$ and $t_b$, which will be either $\mu_{\rm{T}}$ and $\sigma_{\rm{T}}$ for the Gaussian profile or $t_{\rm{start}}$, $t_{\rm{end}}$ for the box profile. For the signal component, we get  
\begin{equation}
    \mathcal{S}(\vec{x}_i, E_i, t_i|\vec{x}_{\rm{S}},\gamma, t_a, t_b) = \mathcal{S}_{\rm{spatial}} \cdot \mathcal{S}_{\rm{energy}} \cdot \mathcal{S}_{\rm{temporal}} = P_\mathrm{PSF}(\vec{x}_i|\vec{x}_{\rm{S}}) \cdot P_{\rm{E}}(E_i|\vec{x}_{\mathrm{S}},\gamma) \cdot P_{\rm{T}}(t_i | t_a, t_b ),
    \label{eq:signal-pdf}
\end{equation}
with $P_\mathrm{PSF}$ as the detector's point-spread-function (PSF) for a source at position $\vec{x}_S$. $P_{\rm{E}}$ defines the probability to observe an event with reconstructed energy $E_i$ originating from $\vec{x}_{\rm{S}}$ and an emission following $E^{-\gamma}$.
%As emphasized in \cite{skyllh:2023icrc}, the energy pdf differs from the pdfs used in internal IceCube analyses.
$P_{\rm{T}}$ describes the probability of observing an event at time $t_i$ if the source only emits during a flare described either by a Gaussian time profile centered at mean $\mu_{\rm{T}}$ with width $\sigma_{\rm{T}}$ or a box profile with constant emission between times $t_{\rm{start}}$ and $t_{\rm{end}}$. 
Similarly, we can write down the background pdf as
\begin{equation}
    \mathcal{B}(\vec{x}_i, E_i, t_i) = \mathcal{B}_{\rm{spatial}} \cdot \mathcal{B}_{\rm{energy}} \cdot \mathcal{B}_{\rm{temporal}} = P(\vec{x}_i) \cdot P(E_i) \cdot P(t_i),
    \label{eq:background-pdf}
\end{equation}
where $P(\vec{x}_i)$ is the known probability of observing a background event at location $\vec{x}_i$, $P(E_i)$ the probability of the event having energy $E_i$, and the time part is assumed to be uniform, i.e. $P(t_i) = \frac{1}{\rm{lifetime}}$. All background pdf parts are independent of signal parameters.

Since we do not know a priori when we expect the source to flare, we first focus on how well the spatial and energy signal and background expectations describe each event. We calculate the ``Signal over Background'' ($S/B$) ratio as
\begin{equation}
    S/B \equiv \frac{\mathcal{S}(\vec{x}_i, E_i|\vec{x}_{\rm{S}},\gamma)}{\mathcal{B}(\vec{x}_i, E_i)} = \frac{\mathcal{S}_{\rm{spatial}} \cdot \mathcal{S}_{\rm{energy}}} {\mathcal{B}_{\rm{spatial}} \cdot \mathcal{B}_{\rm{energy}}} .
    \label{eq:SoB_ratio}
\end{equation}

\begin{figure}
    \centering
    \includegraphics[width=\textwidth]{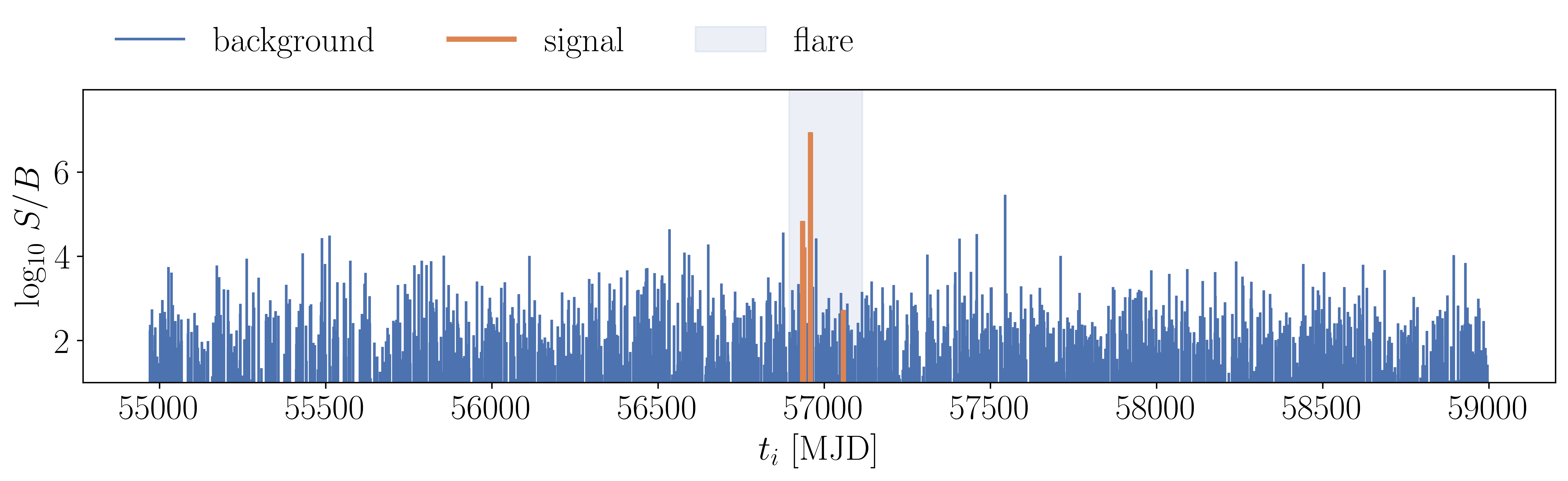}
    \caption{Each event is weighted with its $S/B$ ratio. We highlight the simulated signal events (following an $E^{-2}$ spectrum) in orange. The events are now weighted stronger as signal-like but the flare is still not easily distinguishable.}
    \label{fig:sob_with_signal_highlight}
\end{figure}

Including this time-independent information in our example, we now plot the $S/B$ per event in Fig.~\ref{fig:sob_with_signal_highlight}. We can already see by eye how signal events are up-weighted with respect to background events.

To build the actual distribution of events, including a source, we need to further define a mixture of the signal and background process. Since we do not know the intensity of our source and, therefore, the number of $n_s$ signal events out of the total $N$ observations, we leave $n_s$ as a free parameter. The $\mathcal{S}+\mathcal{B}$ mixture model then becomes $\frac{n_s}{N}\mathcal{S} + (1 - \frac{n_s}{N})\mathcal{B}$. 

The task is now to find estimators for the time parameters of the flare $\hat{t}_a$ and $\hat{t}_b$, the energy parameter $\hat{\gamma}$, and the number of signal events $\hat{n}_s$ (or in general any other parameters involved in the model), that maximizes the likelihood. Since any additional parameter not involving the time can be maximized in an outer loop, i.e. by a standard optimization algorithm, we focus here on the ones describing the time pdf.

The difficulty lies in the complexity of the likelihood space in the time domain. First attempts using standard optimization algorithms were highly dependent on initial guesses and could not reliably correctly identify flare parameters and repeatedly converged to false optima. The only viable option was to perform a "brute force" scan over $t_a$ and $t_b$ \cite{2018Sci...361..147I, abbasi2021search, aartsen2015searches, 2021ApJ...911...67A}, which is computationally extremely costly. In the following section, we present an alternative based on Expectation Maximization.

\subsection{Expectation Maximization}

Expectation Maximization (EM) \cite{10.2307/2984875} is an unsupervised learning algorithm based on Gaussian mixture models. A set of $ K $ Gaussian distributions describes $ N $ observed data points. $ K $ has to be defined in advance and each distribution's mean values and widths are optimized. The Gaussians can be multivariate, hence this approach works for $ M $-dimensional data. The EM algorithm is an iterative procedure, and the general description is \cite{press2007numerical}:

\paragraph{Expectation step (E):}
We calculate the probability $ P(k|i) $ for each data point, $ i $, to belong to a Gaussian distribution $ k $. The estimated parameters are:
\begin{itemize}
    \item $ \mu_k $: the $ K $ means
    \item $ \Sigma_k $: the $ K $ covariance matrices (with dimension $M \times M$)
    \item $ P(k|i)$: the $ K $ probabilities for each data point $ i $ of $ N $, also called the responsibility matrix (the responsibility of component $k$ for data point $i$).
\end{itemize}
$P(k)$ is the probability that a random data point ``belongs'' to Gaussian $k$ or, in different words, $P(k)$ is the fraction of all data points $\vec{y}_i$ originating from $k$. 
The likelihood $ \mathcal{L} $ is the product of the probabilities of observing a data point at its observed value $ \vec{y}_i $

\begin{equation}
\mathcal{L} = \prod_{i}^N P(\vec{y}_i).
\end{equation}
The Gaussian contributions of $P(\vec{y}_i)  $ are
\begin{equation}
P(\vec{y}_i) = \sum_{k}^K \mathcal{N}(\vec{y}_i|\mu_k, \Sigma_k)P(k),
\end{equation}
with $ \mathcal{N}(\vec{y}_i|\mu_k, \Sigma_k)$ as the Gaussian distribution with mean $\mu_k$ and covariance matrix $\Sigma_k$. $ P(k) $ gives the overall weight for component $k$ in the mixture, which can be interpreted as the fraction of all data points that belong to component $ k $.  
The probabilities for each data point $i$ to belong to distribution $k$ are 
\begin{equation}
P(k | i) = \frac{\mathcal{N}(\vec{y}_i | \mu_k, \Sigma_k ) P(k)}{P(\vec{y}_i)}.
\end{equation}

With these equations, it is possible to calculate $ \mathcal{L} $ and the responsibility matrix $P(k|i)$, knowing $ \mu_k, \sigma_k$, and $P(k)$. This is called the expectation step (E-step). 

\paragraph{Maximization step (M):}
The maximization step calculates $ \mu_k, \sigma_k $, and $ P(k) $: 

\begin{equation}\label{eq:maximization_mu}
\hat{\mu}_k = \frac{\sum_i^N P(k|i)\vec{y}_i } {\sum_i^N P(k|i)}, 
\end{equation}
\begin{equation}\label{eq:maximization_mu1}
    \hat{\Sigma}_k =\frac{ \sum_i^N P(k|i) (\vec{y}_i - \hat{\mu}_k) \otimes (\vec{y}_j - \hat{\mu}_k) }{\sum_i^N P(k|i)},
\end{equation}
and thus
\begin{equation}\label{eq:maximization_pk}
\hat{P}(k) = \frac{1}{N} \sum_i^N P(k|i).
\end{equation}
Equations \ref{eq:maximization_mu}, and \ref{eq:maximization_pk} are the maximization step (M-step).

\paragraph{Procedure:} With the E and M steps defined, we follow an iterative procedure:
\begin{enumerate}
    \item Guess starting values for $ \mu_k, \sigma_k$, and $P(k)$.
    \item Repeat: 
    \begin{itemize}
        \item E-step to calculate new $ P(k|i) $, and new $ \mathcal{L} $
        \item M-step to determine new $ \hat{\mu}_k, \hat{\sigma}_k$, and $P(k)$.
    \end{itemize} 
    \item Stop when $ \mathcal{L} $ has converged.
\end{enumerate}

\subsection{Application of EM to the Transient Search}

Let us consider the one-dimensional data from the previous example of a single flare in a time series ($M = 1$).
Considering the defined signal and background hypotheses, the model is a mixture of $K$ Gaussians (signal) and one uniform background distribution. Hence we note $\mu$ as the mean flaring time, $\Sigma$ as the covariance matrix describing the flare width, and $t_i$ as the time event $i$ was detected.
As mentioned in Sec.~\ref{sec:flares}, to describe the data including a flare, we consider a mixture of signal and background components, $\mathcal{S}$ and $\mathcal{B}$. 
The component for the signal with $n_s$ events is 
\begin{equation}
    z_S = \frac{n_s}{N} S(\vec{x}_i, E_i|\vec{x}_{\rm{S}},\gamma) \times \sum_{1}^{K} \mathcal{N}(t_i|\mu_k, \Sigma_k) P(k),
\end{equation}
with $S(\vec{x}_i, E_i|\vec{x}_{\rm{S}},\gamma)$ as the product of the energy and spatial pdfs as in equation \ref{eq:SoB_ratio}, and the temporal pdf consists of a mixture of $K$ Gaussian. 
Similarly, the background component is
\begin{equation}
    z_B = \left( 1 - \frac{n_s}{N} \right) B(\vec{x}_i) \times B_{\rm{temp}}.
\end{equation}
Here, $ B(\vec{x}_i)$ is the product of the spatial and energy background pdfs as in equation \ref{eq:SoB_ratio} and $B_{\rm{temp}} = \frac{1}{\rm{livetime}}$. 

\begin{figure}
    \centering
    \includegraphics[width=\textwidth]{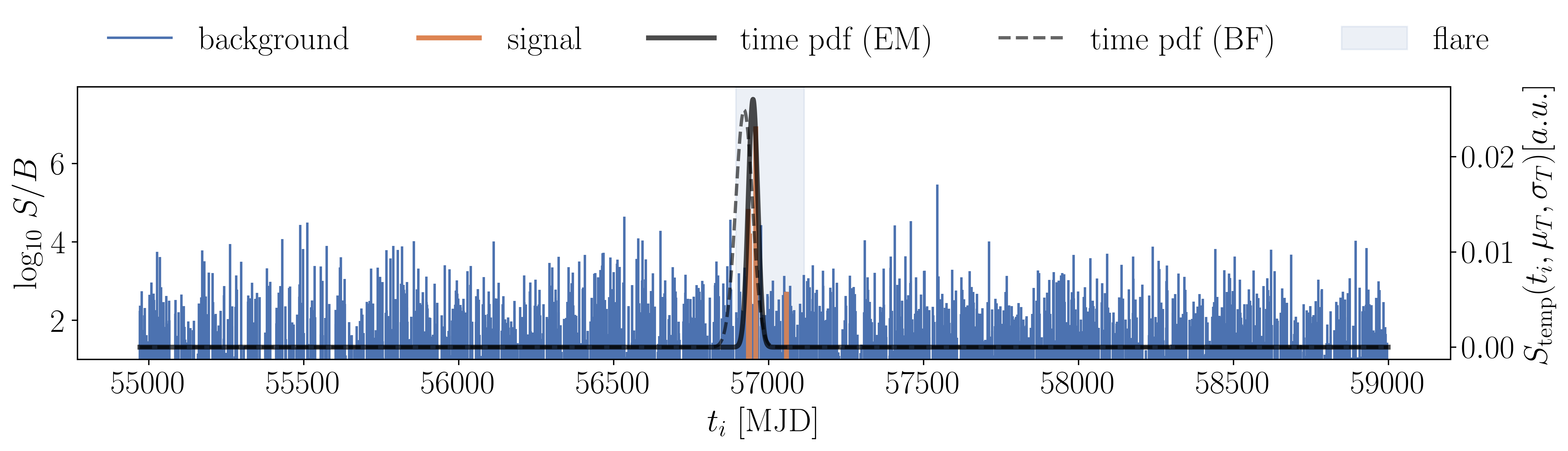}
    \caption{With EM, we find the simulated flare of 5 events within 0.38 seconds. The brute force scan considered possible time intervals with $S/B >= 1$ and took approximately 34 hours to scan the time series of eleven years on a single CPU (Intel Core i7 8th Generation CPU).}
    \label{fig:sob_with_signal_fit}
\end{figure}

For simplicity, let us consider the case for exactly one Gaussian component. i.e. K=1, in the following. Since we added a background component to the mixture, then the correct responsibility matrix becomes
\begin{equation}\label{eq:em_resp_matrix}
    \begin{split}
  P(k=1|i) &= \frac{z_S}{z_S + z_B} = \frac{\frac{n_s}{N} S (\vec{x}_i, E_i|\vec{x}_{\rm{S}},\gamma) \mathcal{N} (t_i | \mu, \Sigma) }
  {\frac{n_s}{N} S (\delta_i, E_i|\delta_{\rm{S}},\gamma) \mathcal{N} (t_i | \mu, \Sigma) + (1 - \frac{n_s}{N}) B (\vec{x}_i) \frac{1}{\rm{livetime}}} \\
    &= \frac{ n_s \ S/B\ \mathcal{N}(t_i | \mu, \Sigma) } { n_s\ S/B\ \mathcal{N}(t_i | \mu, \Sigma) + \frac{N - n_s}{\rm{livetime}}} .
        \end{split}
\end{equation}
In Figure~\ref{fig:sob_with_signal_fit}, we use the above expressions to fit the simulated flares with EM.
The convergence criteria used are $>500$ iterations or no change of the likelihood during 20 iterations.
Comparing the best-fit temporal pdf with the simulated signal, the best-fit temporal pdf described the most significant simulated events. One signal event lies relatively far outside of the resulting Gaussian shape, but this event has a relatively low $S/B$ value.

\begin{figure}
    \begin{minipage}{0.49\textwidth}
    \includegraphics[width=\textwidth]{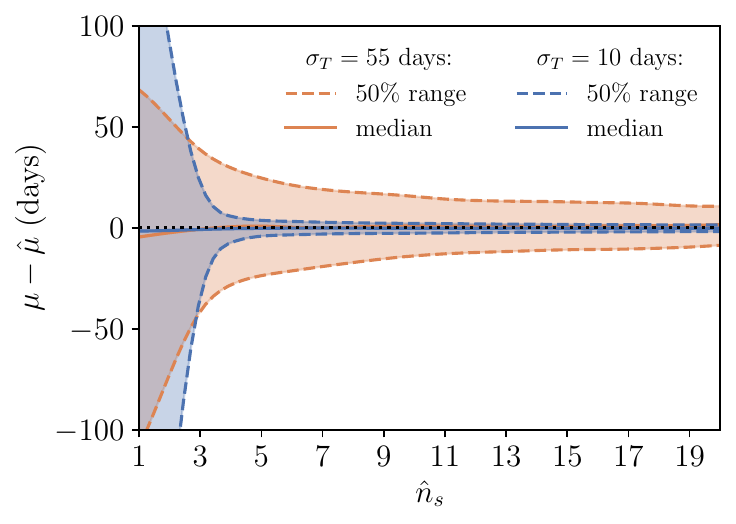}
    \end{minipage}
    \begin{minipage}{0.52\textwidth}
    \includegraphics[width=\textwidth]{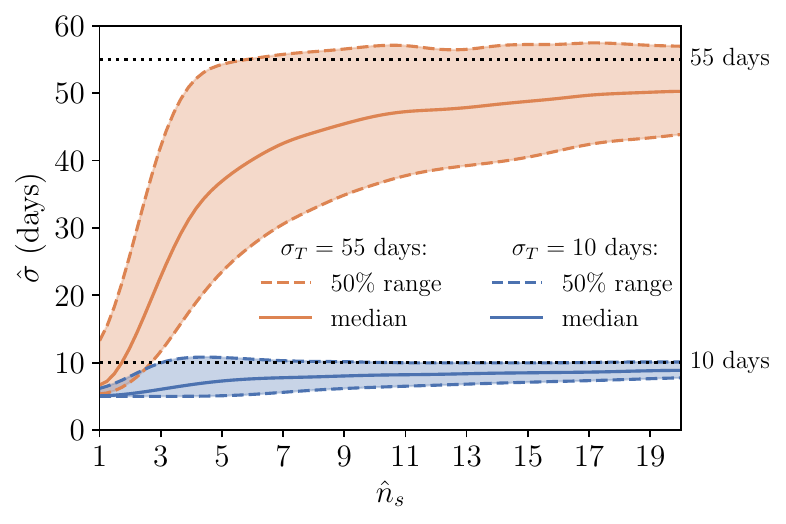}
    \end{minipage}
    \caption{Fit parameter recovery test for two flares with width $\sigma_\mathrm{T} = 10$~days (blue) and $\sigma_\mathrm{T}=55$~days (orange). The solid line shows the median, and the bands show the central 50\% quantiles of the optimized parameter distribution. We show the optimized parameter (deviation) as a function of the number of detected events, $\hat{n}_s$. \textbf{Left:} The deviation of the fitted $\hat{\mu}_\mathrm{T}$ from the actual simulated flare $\mu_\mathrm{T}$ vs. the fitted flare strength. \textbf{Right:} The fitted $\hat{\sigma}_\mathrm{T}$ vs. the fitted flare strength.}
    \label{fig:enter-label}
\end{figure}

We test how well EM recovers a simulated flare for two different time windows of $\sigma_\mathrm{T} = 10$~days and $\sigma_\mathrm{T} = 55$~days, see Fig.~\ref{fig:enter-label}. In both cases, we assume a point source at a given position. The quantities $\hat{\mu}, \hat{\sigma}$, and $\hat{n}_s$ are the results of EM optimization as specified in equation \ref{eq:em_resp_matrix}, whereas $\mu, \sigma$, and $n_s$ refer to the true parameters. The true mean flaring time, $\mu_\mathrm{T}$, is recovered well already for few detected signal events, $\hat{n}_s$, on average $\gtrsim 3$. The flare duration (or width) $\sigma_\mathrm{T}$ is underestimated for small numbers of signal events. In the most extreme case of only a single signal event, the width approaches the minimal allowed value (here 5\, days), as is expected from the sample variance of one event being zero. For stronger signals, the fitted $\hat{\sigma}_\mathrm{T}$ approaches the true value. The number of events needed to identify the flare depends on the actual signal properties and the assumed likelihood for calculating $S/B$. In this example case, the simulated signal follows a power-law energy spectrum $\propto E^{-2}$, whereas the background events are best described by a power-law energy spectrum $\propto E^{-3.7}$. A source emitting events more similar to the background (for example, $\propto E^{-3}$) will require a stronger flare for identification than a source emitting a very distinct signal (for example, $\propto E^{-1}$) from the background. Hence, the number of events necessary for flare identification varies depending on the signal and background realizations and how well the likelihood describes the data.

\begin{figure}
    \centering
    \includegraphics[width=\textwidth]{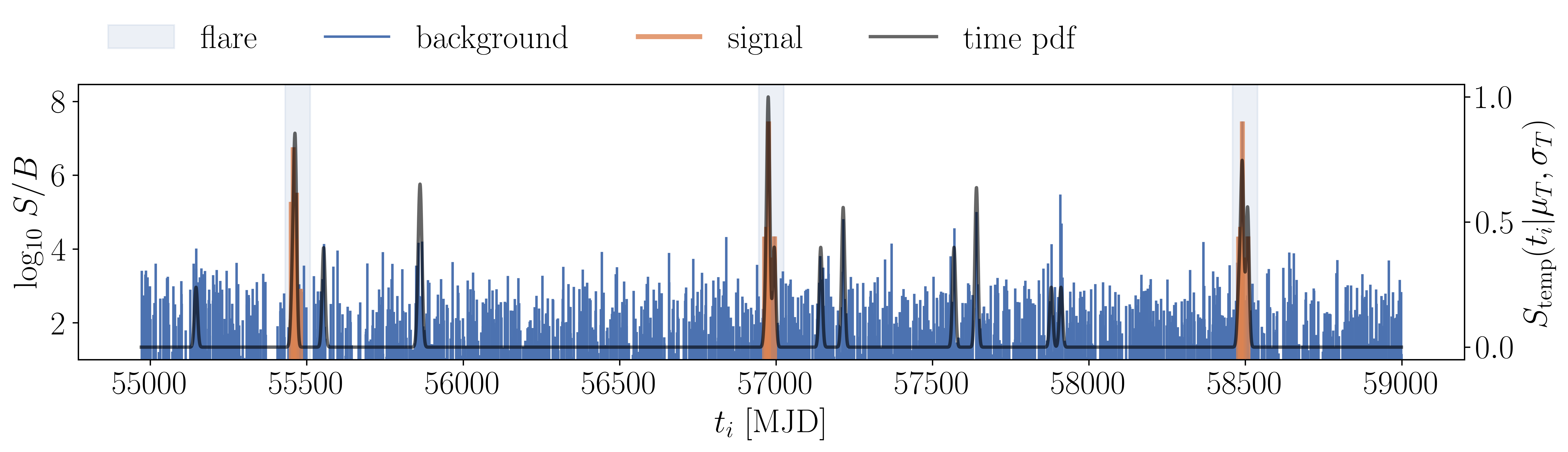}
    \caption{Three simulated flares with each 5 events and a spread of $\sigma_{\rm{T}} = 20$~days. With EM we fit multiple ($K=100$) distributions. Most are fit to a weight of 0. We recover the simulated flares and also find some less significant background fluctuations.}
    \label{fig:multiflare}
\end{figure}

Our algorithm is easily expandable to find multiple flares. We set $K > 1$ for multiple flares and evaluate the mixture model in analogy. In the example shown in Figure~\ref{fig:multiflare}, we simulate three flares with five signal events each and a width of $\sigma_{\rm{T}} = 20$~days. We set $K=100$ since, as for real data, we would not know how many flares we expect. The best number for $K$ to be used is analysis dependent and best determined on simulated data; we found that $K=100$ works well in our application. Most flares are fit to a weight $P(k)=0$, so they do not contribute. Our EM algorithm correctly finds the simulated flares and fits a few background fluctuations as less significant flares. 

When comparing a constant emission between $t_{\rm{start}}$ and $t_{\rm{end}}$ with a Gaussian-shaped flare of similar duration, EM identifies both signals equally well, even when the simulated emission is box-shaped and does not follow the fitted Gaussian pdf \citep{2022icrc.confE.940K}. 

\subsection{Comparison Between EM and Brute-Force Search}

We compare the performance of the EM algorithm with the conventional brute-force approach. In the latter, we assume a time-dependent signal between $t_{\rm{start}}$ and $t_{\rm{end}}$, selecting these time boundaries based on event detection times surpassing a specified $S/B$ threshold. We optimize the log-likelihood ratio for all potential intervals within the minimum duration of 5 days and maximum duration of 300 days. Various $S/B$ thresholds are tested, with higher values reducing the number of possible flare intervals. Only for the brute-force method $S/B$ thresholds are applied, whereas for EM, all data are included without any cuts.

\begin{figure}[h]
    \centering
    \includegraphics[width=\textwidth]{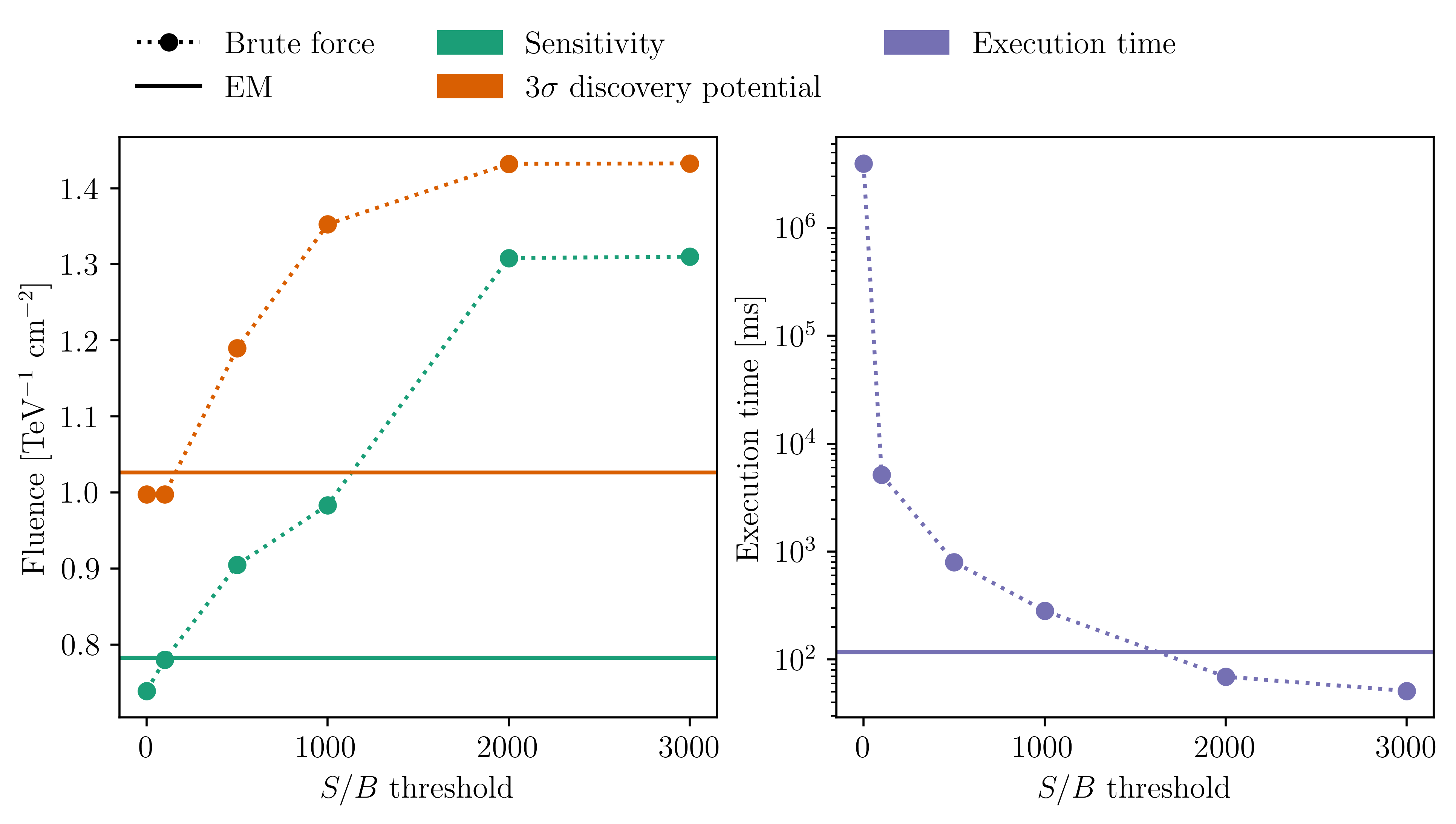}
    \caption{Comparison of expectation maximization (no $S/B$ cuts) with the brute-force approach for different $S/B$ thresholds for a time series spanning one year. \textbf{Left:} The detectable fluxes integrated over the respective time window (fluences). We show the sensitivity in green and the discovery potential in orange for EM and the brute force approach. With EM, we detect fluences (on the y-axis) similar to a brute force approach applying a threshold of $S/B \sim 100$. \textbf{Right:} The computation time (y-axis) decreases with increasing $S/B$ threshold. EM finds the flares faster than a brute-force approach with a threshold of $S/B > 1000$ while being as sensitive as a brute-force search with a threshold of $S/B \sim 100$. }
    \label{fig:em_sob_comparison}
\end{figure}

For this comparison, the total uniform background data spans one year. We simulate a flare lasting 110 days and vary the number of signal events within the flare to assess the identifiable source flux within the time window. Since, in general, the detectable flux depends on the width of the flare, we integrate the fluxes over the respective time windows, resulting in the fluence. The left panel of Fig.~\ref{fig:em_sob_comparison} displays the sensitivity fluence (green) and $3\sigma$ discovery potential fluence (orange). The discovery potential is the fluence we could detect with 3$\sigma$ significance on a 50\% confidence level. In case of a non-detection, the sensitivity is the 90\% confidence level upper limit fluence we could set \cite{AHRENS2004507}.
The right panel of Fig.~\ref{fig:em_sob_comparison} shows the computational times for brute-force and EM approaches. While the brute-force method can detect weaker sources when applying very small (or no) $S/B$ thresholds compared to EM, the computational times for small thresholds are orders of magnitude ($\sim 10^4$) larger than for EM. EM offers comparable sensitivity to a brute-force approach with $S/B \sim 100$ but with significantly faster computation.

In previous brute-force searches, analyses employing small $S/B = 1$ thresholds were either limited to few years of neutrino data \cite{aartsen2015searches, 2021ApJ...911...67A} or focused solely on a single position \cite{2018Sci...361..147I}, making scalability to larger datasets challenging. Another search \cite{abbasi2021search} utilized thresholds $S/B \gg 100$. However, with EM, we introduce a computationally efficient algorithm capable of incorporating all available information while achieving greater sensitivity than the alternative approach of employing large $S/B$ thresholds. In ref. \cite{alert_followup_paper}, our EM approach is applied for the first time to IceCube data, which allowed for a comprehensive search for transient emissions from many IceCube high-energy alert positions. As demonstrated, EM is also scalable to multiple flares and adaptable to growing datasets, while the increased search space complexity is prohibitive to using brute force.

\section{Example Application: IceCube Neutrino Flare from TXS~0506+056}\label{sec:TXS}
We apply the above-presented algorithm on the example case of the neutrino flare from TXS~0506+056 \cite{2018Sci...361..147I}. The background assumption is a spatially and temporally uniform background only depending on the detector's effective area. Our signal assumption is a Rayleigh distribution centered at TXS~0506+056 as the spatial pdf, an energy distribution following a power-law ($\propto E^{-\gamma}$), and a temporal pdf of a Gaussian. 

To calculate the $S/B$ ratio, we must assume a source spectral index, $\gamma$, and since we do not want to favor a specific spectral index, we calculate $S/B$ for different $\gamma$ values and run EM on the resulting, different $S/B$ ratios. For each $S/B$ ratio, we determine the best temporal pdf and use it to perform a hypothesis test optimizing $n_{\rm{S}}$ and $\gamma$. Ultimately, we chose the most significant result as our best-fit parameters. The procedure step-by-step is to:

\begin{samepage}
\begin{enumerate} %TODO indent different 'layers'
    \item Select a position $\vec{x}$. 
    \item Calculate $S/B$ for a specific spectral index $\gamma$.
    \item Run EM and determine the best fit $\hat{\mu}_{\rm{T}}$ and $\hat{\sigma}_{\rm{T}}$.
    \item Use $\hat{\mu}_{\rm{T}}$ and $\hat{\sigma}_{\rm{T}}$ for fixing the temporal signal pdf. Run a subsequent likelihood ratio test optimizing for $n_{\rm{S}}$ and $\gamma$.
    \item Repeat steps 2. to 4. for different spectral indices (in our case, in the range [1.5, 4] with steps of 0.2).
    \item Choose the result yielding the best likelihood ratio.
\end{enumerate}
\end{samepage}

\begin{figure}
    \centering
    \includegraphics[width=\textwidth]{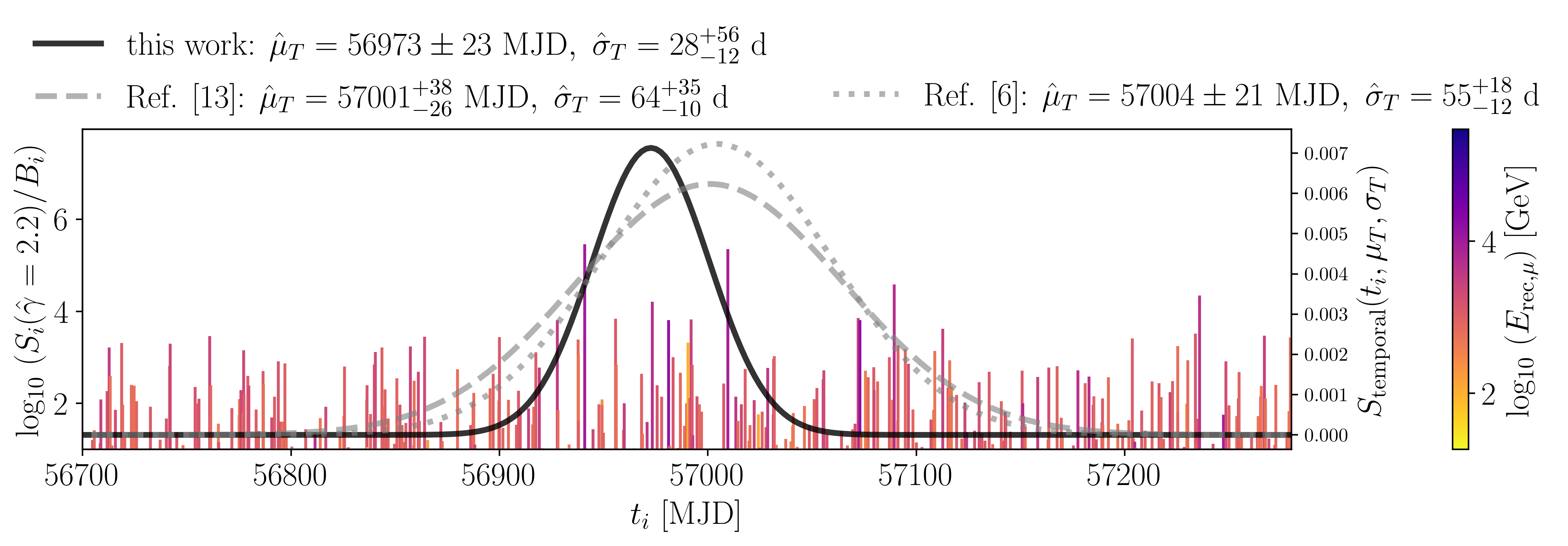}
    \caption{The S/B weights for events near TXS~0506+056 vs. detection time. We show the best-fit neutrino flare of this work (solid black line) compared to the best-fit flares of ref. \cite{2018Sci...361..147I} and ref. \cite{alert_followup_paper}.}
    \label{fig:txs_flare_sob}
\end{figure}

Following this procedure, we apply EM on IceCube public data at the position of TXS~0506+056 to reproduce the analysis in refs.~\cite{2018Sci...361..147I, abbasi2021search, alert_followup_paper}. Similar to ref. \cite{skyllh:2023icrc}, we identify the neutrino flare at a mean time of $\hat{\mu}_{\rm{T}} = 56973 \pm 23$~(MJD) with a width of $\hat{\sigma}_{\rm{T}}= 28 ^{+56}_{-12}$~days. The best-fit parameter of the mean number of neutrinos is $\hat{n}_{\rm{S}} = 7 ^{+6}_{-5}$ and for the spectral index we get $\hat{\gamma} = 2.2^{+0.5}_{-0.4}$. The results are compatible with published IceCube results in refs. \cite{2018Sci...361..147I, abbasi2021search, alert_followup_paper} (see Fig.~\ref{fig:txs_flare_sob}), which use more precise energy pdfs based on unpublished Monte Carlo data \cite{skyllh:2023icrc}. This leads to different signal weights for the public data analysis than in the internal IceCube analyses and influences the flare's best-fit parameters. 

\section{Conclusion}
We presented a new method for fitting transient emission using Expectation Maximization (EM). This approach also successfully identifies weak signals in heavily background-dominated data, such as when studying astrophysical neutrino emission. Our method significantly reduces the computational burden required to identify flares compared to conventional techniques, achieving a speedup of a factor of over $10^4$ on a single CPU.
EM is based on a Gaussian mixture and offers the flexibility to fit different models. It seamlessly transitions from analyzing single flares to handling multiple flares and is compatible with multi-dimensional data. 
We showcase the effectiveness of EM by applying it to a simulated time series of detected events. Furthermore, we demonstrate its ability by employing EM to identify the IceCube neutrino flare of TXS~0506+056, yielding best-fit parameters consistent with previously published results.
By providing a fast and sensitive method for identifying transients, our approach enables the extension of searches to a broader range of source candidates, including highly variable objects like active galactic nuclei, or even to encompass the whole sky for an ever-increasing amount of data.

\section*{Acknowledgements}

This project was supported by the Deutsche Forschungsgemeinschaft (DFG, German Research Foundation) under Germany's Excellence Strategy – EXC-2094 – 390783311, and the Sonderforschungsbereich (Collaborative Research Center) SFB1258 ‘Neutrinos and Dark Matter in Astro- and Particle Physics’. 

% \acknowledgments

% This is the most common positions for acknowledgments. A macro is
% available to maintain the same layout and spelling of the heading.

% \paragraph{Note added.} This is also a good position for notes added
% after the paper has been written.

% Bibliography

%% [A] Recommended: using JHEP.bst file
\bibliographystyle{JHEP}
\bibliography{biblio.bib}

%% or
%% [B] Manual formatting (see below)
%% (i) We suggest to always provide author, title and journal data or doi:
%% in short all the informations that clearly identify a document.
%% (ii) please avoid comments such as "For a review'', "For some examples",
%% "and references therein" or move them in the text. In general, please leave only references in the bibliography and move all
%% accessory text in footnotes.
%% (iii) please have only one work for each \bibitem.

% \begin{thebibliography}{99}

% \bibitem{a}
% Author,
% \emph{Title},
% \emph{J. Abbrev.} {\bf vol} (year) pg.

% \bibitem{b}
% Author,
% \emph{Title},
% arxiv:1234.5678.

% \bibitem{c}
% Author,
% \emph{Title},
% Publisher (year).

% \end{thebibliography}

\end{document}